\documentclass[submission,copyright,creativecommons]{eptcs}
\providecommand{\event}{FMAS 2025} 
\ifpdf
  \usepackage{underscore}         
  \usepackage[T1]{fontenc}        
\else
  \usepackage{breakurl}           
\fi

\usepackage{wrapfig}

\usepackage{cite}
\usepackage{graphicx}
\usepackage{amsmath,amssymb,amsfonts}
\usepackage{algorithmic}
\usepackage{textcomp}
\usepackage{xcolor}
\usepackage{multirow}
\usepackage{csquotes}
\usepackage[inline]{enumitem}
\usepackage{xspace}
\usepackage{sty/commands}
\usepackage{xcolor}
\usepackage{hyperref}
\hypersetup{
    colorlinks,
    linkcolor={red!50!black},
    citecolor={blue!50!black},
    urlcolor={blue!80!black}
}

\def\BibTeX{{\rm B\kern-.05em{\sc i\kern-.025em b}\kern-.08em
    T\kern-.1667em\lower.7ex\hbox{E}\kern-.125emX}}

\title{Model Learning for Adjusting the Level of Automation in HCPS}
\author{
Mehrnoush Hajnorouzi \qquad \quad Astrid Rakow
\institute{German Aerospace Center (DLR) e.V.\\
Oldenburg, Germany}
\email{mehrnoush.hajnorouzi@dlr.de \qquad astrid.rakow@dlr.de}
\and
Martin Fränzle
\institute{
Carl von Ossietzky Universität Oldenburg\\ 
Oldenburg, Germany}
\email{martin.fraenzle@uni-oldenburg.de}
}

\def\titlerunning{Adjusting level of automation in \HCPS}
\def\authorrunning{M. Hajnorouzi, A. Rakow \& M. Fränzle}

\begin{document}
\maketitle

\begin{abstract}
The steadily increasing level of automation in human-centred systems demands rigorous design methods for analysing and controlling interactions between humans and automated components, especially in safety-critical applications. The variability of human behaviour poses particular challenges for formal verification and synthesis. 
We present a model-based framework that enables design-time exploration of safe shared-control strategies in human–automation systems. The approach combines active automata learning—to derive coarse, finite-state abstractions of human behaviour from simulations—with game-theoretic reactive synthesis to determine whether a controller can guarantee safety when interacting with these models. If no such strategy exists, the framework supports iterative refinement of the human model or adjustment of the automation's controllable actions. 
A driving case study, integrating automata learning with reactive synthesis in \textsc{Uppaal}, illustrates the applicability of the framework on a simplified driving scenario and its potential for analysing shared-control strategies in human-centred cyber-physical systems.
\end{abstract}

\section{Introduction}
Formal verification and synthesis methods have achieved significant success for automated systems, where the dynamics and control logic of the system can be precisely modelled. However, this paper focuses on human-centred cyber-physical systems (\HCPS) that operate under a shared-control paradigm~\cite{sheridan78}. In such systems, humans and automation jointly contribute to task execution 
(\eg in advanced driver-assistance systems \cite{Marcano20}, robot-assisted surgery \cite{Scheikl21}, or flight-control systems \cite{Goodrich06}), and the level of automation must be dynamically adapted according to the evolving system state and the human’s behaviour. This poses a major modelling challenge, since the \enquote{human component} must also be represented in a form that can be analysed by formal methods. 

Many shared-control \HCPS are safety-critical and must satisfy stringent safety and performance requirements. A key challenge lies in managing the dynamic and bidirectional interaction between humans and automation, ensuring coordinated actions that maintain safety and efficiency despite behavioural uncertainty. This calls for control strategies that not only optimise the automation’s capabilities but also account for the inherent unpredictability of human cognition and decision-making. Because human cognition is variable, context-dependent, and only partially observable, constructing models that support formal safety guarantees remains a major challenge~\cite{FMHAI2013,Bolton2010,Webster2019}.

This work aims to support the design of safety-critical shared-control \HCPS. In their design, two central questions arise: \emph{(Q1) \enquote{What does the automation need to know about the human component to interact appropriately?}} and \emph{(Q2) \enquote{What must the automation do in order to interact appropriately?}}. Naturally, these questions are interrelated. \expl{Consider a driver-support system that intervenes only to prevent collisions. If the automation can detect an inattentive driver early, a slight alert might suffice to avoid an imminent collision; but if the driver fails to respond, emergency braking must be triggered to ensure safety.} We provide a method that aligns the automation’s capabilities with the controller’s knowledge of the human.

\begin{figure}
    \centering
    \includegraphics[width=0.55\textwidth]{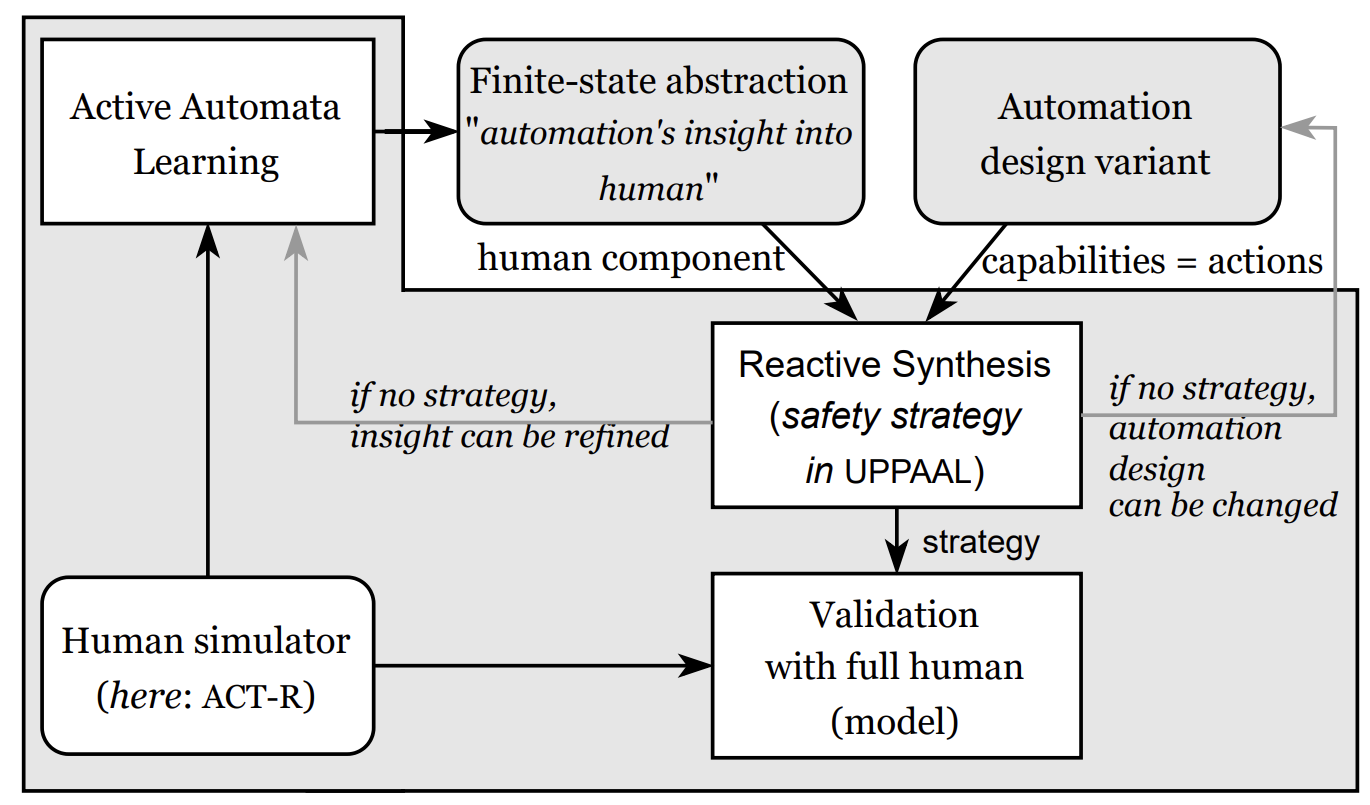}
    \caption{Overview of the approach: Does the insight into the human and the automation capabilities suffice to implement a safe shared-control \HCPS?}
    \label{fig:approach}
\end{figure}

Our approach (\cf Fig. \ref{fig:approach}) supports this design process through a design-time analysis framework. 
A human simulation engine—in this paper, the cognitive architecture {\smaller{ACT-R}}—generates observable human behaviour, which is abstracted via active automata learning into a finite-state model representing the automation’s operational view of the human.
Given this abstraction and a specific automation design, we apply game-theoretic reactive synthesis to determine whether the automation can guarantee safety and fulfill its mission goals. 
If no winning strategy is found, we either refine the human model or adjust the automation’s capabilities. 
Since the strategy is synthesised with respect to the learned human abstraction and scenario mode, it is subsequently validated through co-simulation with the full cognitive model. 
When synthesis and validation both succeed, the derived models specify the level of human state observability and automation capability required for implementation.

Formal methods have been applied to human–automation interaction~\cite{Bolton2010,Webster2019}, yet mostly rely on static or normative human models, limiting  applicability in dynamic or adaptive contexts. 
Our framework addresses this limitation by integrating model learning and reactive synthesis to enable reasoning about shared-control with behaviourally derived human abstractions.
We demonstrate the approach in a simplified driving scenario that combines active automata learning from a cognitive architecture with reactive synthesis in \textsc{Uppaal}, to analyse safety properties under varying human behaviour.
Rather than focusing on isolated control actions, this work addresses the formal analysis of continuous interaction over unbounded time horizons \cite{Anderson24}, characterised by dynamically evolving event sequences arising from the coupled dynamics of the human, automation, and environment.

The contributions of this paper are threefold:
\begin{enumerate*}[label=(\arabic*)]
    \item a design-time framework that integrates model learning and reactive synthesis for analysing shared-control in \HCPS;
    \item a method for deriving finite-state human abstractions through interaction with a cognitive simulation; and
    \item a preliminary evaluation in a driving scenario demonstrating the feasibility of synthesising safety-preserving automation strategies.
\end{enumerate*}

\textbf{Structure of the article.}
Section~\ref{sec:prelim} introduces the key design considerations for \HCPS, 
Section~\ref{sec:method} details the proposed framework.
Section~\ref{sec:caseStudy} presents a brief case study. 
Related work is reviewed in Section~\ref{sec:RW}.
Section~\ref{sec:discussion} summarises the main findings, discusses limitations and outlines directions for future research.
Finally, Section~\ref{sec:conclu} concludes the paper.

\section{Preliminaries}
\label{sec:prelim}

Figure~\ref{fig:approach} provides an overview of the proposed framework for analysing shared-control in \HCPS. 
To facilitate understanding of the subsequent sections, this section introduces the main concepts and notations required to comprehend the overall approach. 
We briefly recall the notions of human–automation interaction, finite-state models, and reactive synthesis, as well as the concept of adjusting the level of automation, which together constitute the methodological foundation of the framework.

\subsection{Human--Automation Interaction}
\label{sec:human}
Designing \HCPS requires explicit consideration of the bidirectional feedback between automation and human operators: the automation's actions influence user behaviour, and human actions in turn affect the automation. 
To represent the human component of an \HCPS, we employ computational models of human decision-making based on cognitive architectures. 
Architectures such as \ACTR~\cite{Anderson04} and \CASC~\cite{ludtke2009} provide general frameworks grounded in psychological theory~\cite{newell90,anderson98} and have been validated empirically across diverse human-factors domains (\eg \cite{Salvucci01, ludtke11validation}).
They generate discrete, time-stamped behavioural traces that can serve as observable input--output data for learning and analysis.

The \ACTR\ architecture models human behaviour as a perception--cognition--action loop operating over a sequence of cognitive cycles (typically 50\,ms each). 
At every cycle, perceptual modules process sensory input from the simulated environment; the central procedural system selects a production rule that determines the next cognitive or motor action; and the resulting action updates both the environment and the internal state. 
Adjustable parameters such as memory decay, learning rate, and stochastic noise govern inter-individual variability and non-deterministic decision patterns. 
Consequently, \ACTR\ produces rich but discrete behavioural sequences that can be interpreted as a mapping between environmental stimuli and human responses.

In the context of this work, the simulated interactions yield traces of human behaviour that serve as the empirical basis for model inference. 
These traces are abstracted by the active automata-learning procedure into a finite-state representation of the behaviour that the automation \emph{assumes} or \emph{expects} from the human component during interaction. 
We refer to this internal predictive model simply as the abstract \emph{human model} (\HM). 
The \HM thus embodies the automation’s working hypothesis about human responses within the task context and forms the formal interface between cognitive simulation and reactive synthesis, enabling systematic exploration of shared-control strategies.

\subsection{Reactive Strategy Synthesis and Safety Games}
\label{secReactiveS}
The objective of reactive synthesis is to automatically construct strategies that ensure system objectives while adapting to variations in human behaviour, rather than to pre-compute fixed action sequences. 
The interaction between the automation and the human is modelled as a finite-state (timed) two-player game, where the automation aims to maintain system safety and achieve mission goals despite uncertain or variable human actions. 
Such games are commonly referred to as \emph{safety games}, in which a \emph{winning strategy}~\cite{Alur04} for the automation guarantees that no unsafe state can be reached under any admissible human behaviour. 
Desired system properties are expressed in temporal logics such as LTL or TCTL. 
In contrast to the traditional \emph{implement-then-verify} workflow, reactive synthesis produces a \emph{correct-by-construction} controller directly from these formal specifications.

Formally, the system is regarded as a discrete-event system~\cite{Cassandras10}, whose behaviour is represented as a temporally ordered sequence of events. 
Finite automata and their ($\omega$-)regular languages provide a natural formalism for representing and analysing such behaviours. 
Cognitive architectures such as \ACTR, however, are not expressed in an automata-based formalism.
To obtain a representation suitable for synthesis, we employ \emph{active automata learning}.

Active learning techniques based on Angluin’s L* algorithm~\cite{angluin87} iteratively query the system under learning (\SUL), construct hypotheses, and refine them using counterexamples until convergence to a finite automaton that approximates the target behaviour. 
This approach is particularly suitable for modelling human behaviour: data obtained through passive observation are typically incomplete due to behavioural variability and context dependence, and rare yet safety-critical actions may be absent from logs. 
Active learning mitigates these limitations by interactively exploring the behavioural space and refining the hypothesis through counterexample analysis.

Active automata learning has seen numerous extensions in recent years, including optimised algorithms such as \TTT~\cite{IsbernerTTT14} and \NL~\cite{Bollig09} that reduce the number of required queries, as well as probabilistic and non-deterministic variants~\cite{aalpy22,vaadrager2017,Pferscher20} aimed at modelling uncertainty and variability in system behaviour. The development of tools such as \textsc{AALpy}~\cite{aalpy22} and continued research on applying and extending active model learning~\cite{Muskardin24,vonBerg25} underline its ongoing relevance in both theory and practice.
Its simplicity, transparency, and established convergence guarantees make it a natural choice for integration with cognitive simulations, while more efficient variants can be adopted within the same framework without conceptual modification.

\subsection{Levels of Automation and Shared-Control}
\label{sec:automationlevels}
The degree of automation in human–machine systems varies depending on how control authority and decision-making are distributed between the human operator and the automated components.  
The widely adopted taxonomy \SAE{} ~\cite{SAE21}, defines six levels of vehicle automation, ranging from level~0 (no automation) to level~5 (full automation).  
At lower levels, automation offers assistance functions such as adaptive cruise control; intermediate levels enable partial automation that jointly controls acceleration and steering while requiring continuous human supervision; and higher levels allow conditional automation that manages all driving tasks within defined scenarios, returning control to the human when intervention is required.
Across this spectrum, human involvement ranges from hands-on to mind-off interaction.

For such systems, particularly those operating at intermediate levels, the \emph{shared-control} principle becomes essential: both human and automation contribute to task execution, dynamically balancing authority to ensure safety and performance.  
The automation must recognise when to intervene and when to defer to the human operator, adapting its behaviour to the situational context and the human’s state.
In our framework, levels of automation are instantiated as design variants, each defined by its controllable action set available to the automation.
Through reactive synthesis, we evaluate whether a given variant can satisfy the control objectives while maintaining safety and accommodating the human behaviour.  
This enables systematic comparison of automation levels within a unified formal framework.

\section{Iterative Model Learning and Control Synthesis}
\label{sec:method}
\normalsize
Our approach, illustrated in Fig. \ref{fig:approach}, integrates model learning with reactive strategy synthesis to enable adaptive automation in \HCPS. 
The interaction between automation and human is formalised as a two-player game and control strategies are synthesised to guarantee mission objectives under all admissible human behaviours.  

This section presents the methodological core: the construction of finite-state human abstractions and the subsequent synthesis of control strategies.  
Active automata learning derives a finite-state \emph{abstract human model} of the cognitive architecture, representing the information available to the automation about the human. 
The learned model, composed with the environment and the formalised objectives, serves as input to game-theoretic synthesis. 
The process proceeds in three main stages.
\begin{enumerate}
    \item \textbf{Model learning}: construct a finite-state abstraction of the cognitive architecture within its operational context, capturing the information that the automation obtains about the human.
    \item \textbf{Game construction}: combine the learned human abstraction with the environment and the control objectives, encoding them as winning conditions.
    \item \textbf{Synthesis}: compute a winning strategy for the automation that satisfies the objectives against admissible human behaviours.
\end{enumerate}

\subsection{Generating the Abstract Human Model}
\label{sec:GenAbstraction}
Shared-control automation must adapt to variability and limitations of human cognition--- intervening when human state may lead into unsafe situations and remaining otherwise unobtrusive.
To formalise these conditions, we define abstractions of the human–machine interaction at an appropriate level of granularity and learn corresponding finite-state representations from cognitive-architecture simulations.
The resulting automaton represents the automation’s predictive model of human’s behaviour within the task context.

\begin{figure}[t!]
    \centering
    \includegraphics[width=.96\textwidth]{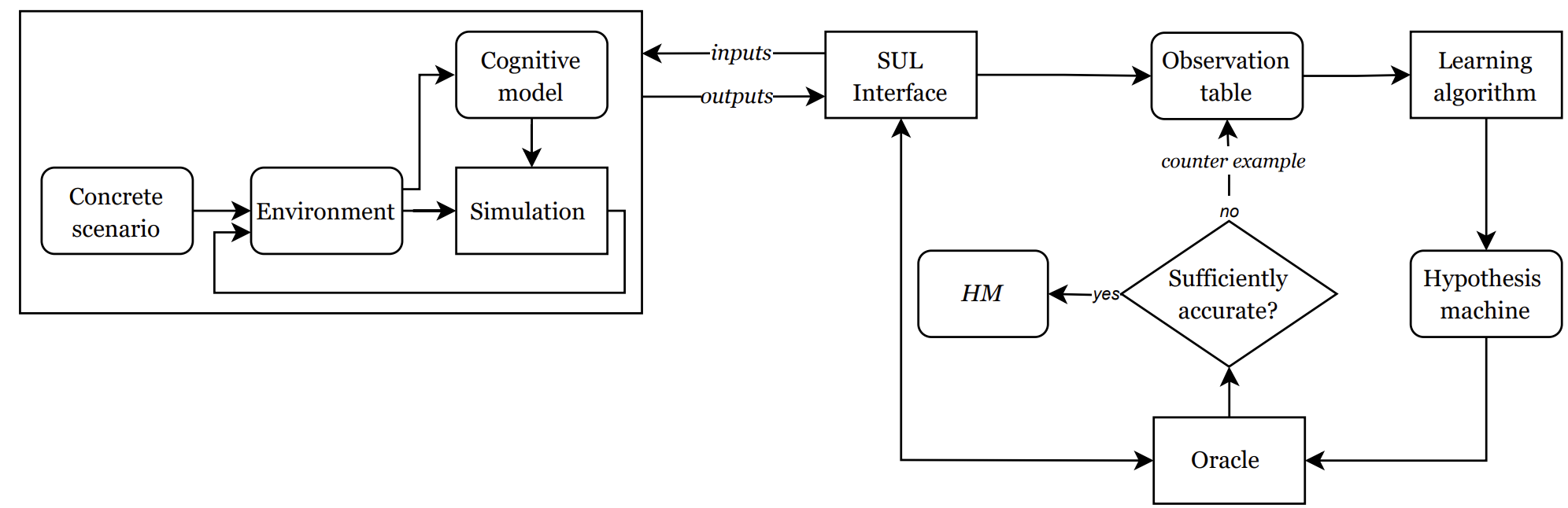}
    \caption{Learning behaviour abstraction automaton (\HM) from simulation of cognitive model.}
    \label{fig:LAHM2}
\end{figure}

The cognitive architectures are inherently state-based but structurally complex and concurrent. 
We use \emph{finite game graphs}~\cite{Thomas95} as the formal basis.  
The abstraction is obtained via \emph{active automata learning} based on Angluin’s L* algorithm~\cite{angluin87}.  
Learning begins from a coarse abstraction--—formed over discrete observation intervals—--and is refined iteratively until sufficient behavioural insight is achieved for the target scenario. 
Each iteration simulates the cognitive model within the task environment, records input–output traces, and updates the learned automaton~(\HM).  This process yields a simplified yet behaviourally representative depiction of the human component (see Fig. \ref{fig:LAHM2}).

The selection of observable parameters follows a pragmatic principle that prioritises variables with measurable correlation to the cognitive state.
Typical examples include perceptual indicators such as facial expressions or gaze direction, which can serve as estimators of mental workload or attention focus (\cf \cite{Gorin24,Diarra25}).
This facilitates the integration of the approach within application-specific domains.
The key criteria are
\begin{enumerate*}[label=(\roman*)]
    \item observability of parameters and
    \item inference effectiveness for the underlying cognitive state.
\end{enumerate*}

Active automata learning incrementally constructs a hypothesis automaton by querying a \emph{system under learning} (\SUL) and refining the hypothesis based on counterexamples. 
The learner maintains an \emph{observation table} 
comprising input prefixes and suffixes, 
and their observed outputs.
For the table to define a valid hypothesis, it must satisfy closure and consistency conditions, ensuring that equivalent prefixes denote the same abstract state.
Once these conditions are met, the learner constructs a hypothesis automaton whose states correspond to equivalence classes of prefixes exhibiting identical output behaviour over all suffixes.
Subsequently, the learner issues equivalence queries to assess behavioural conformance with the
\SUL. In practice, equivalence is approximated through randomised testing, whereby a large set of input sequences is applied to both the hypothesis and the \SUL. 
Any divergence yields a counterexample, which is incorporated into the observation table to refine the hypothesis. 
The iterative cycle of construction, testing, and refinement continues until no counterexamples are identified within the defined bounds.

We note that the expressive power of cognitive models generally exceeds that of regular languages; exact learning is therefore infeasible.  The resulting automaton approximates the behavioural language of the cognitive model, analogous to abstractions learned for recurrent neural networks in related work~\cite{Bollig22}. 
Empirical validation across multiple cognitive architectures can be used to assess the
adequacy of this finite-state approximation for the intended synthesis tasks.
The proposed framework is not restricted to a specific cognitive architecture, such as \ACTR.
Different architectures emphasise distinct aspects of human cognition---for instance, emotional–behavioural interaction in \smaller{MAMID} \cite{Hudlicka02}---and may be selectively employed depending on the application focus.
Combining or comparing multiple architectures further supports cross-validation of the learned abstractions and enhances the robustness of the resulting automation design.
The learned abstraction \HM thus serves as the finite-state human behaviour within the subsequent game-theoretic synthesis stage.

\subsection{Game Graph Construction and Synthesis}
Let \CPS be a design variant of the automation defined by its controllable action set.  
The learned human model~\HM{} is combined with the cyber-physical components (\CPS) and the environment models to form the \HCPS game arena.  The arena is represented as a \emph{timed game automaton} (TGA) \cite{Maler95} \smaller{$\mathcal{A} = \langle L, l_{0}, \Sigma_c, \Sigma_u, X, Inv, E \rangle$},
where $L$ denotes the set of locations, $l_0$ the initial location, $X$ the set of clocks, and $Inv$ assigns invariants to locations.
The transitions $E$ are labelled with actions from the disjoint sets of controllable actions~$\Sigma_c$ (automation) and uncontrollable actions~$\Sigma_u$ (human and environment).  
The synchronous product $\mathcal{A} = HM \parallel CPS$ defines the joint behaviour, interleaving independent actions and synchronising on shared ones.

During execution, each player chooses an action and an associated delay $t \in \mathbb{R}_{\ge 0}$. The opponent may pre-empt by selecting an enabled action with a shorter delay~$t' \le t$.  
The resulting interleaving defines the plays of the game.
In the case study configuration, a fixed cognitive-cycle delay of 50 ms was adopted, corresponding to the standard temporal resolution in cognitive modelling.

The synthesis problem is to construct a control strategy that selects actions in $\Sigma_c$ such that the control objective $\texttt{C}$ is satisfied for all behaviours in $\Sigma_u$.  Control objectives are expressed as winning conditions in a fragment of Timed Computation Tree Logic (TCTL)~\cite{Bouyer17}, typically including:
\begin{itemize}
    \item \emph{Safety}: undesirable states are never reached (\smaller{$\forall \square \lnot bad$});
    \item \emph{Reachability}: desirable states are eventually reached (\smaller{$\forall \lozenge goal$});
    \item \emph{Response}: whenever a trigger holds, a response eventually follows (\smaller{$\forall \square (trigger \Rightarrow \lozenge response)$}).
\end{itemize}
These specifications are compiled into acceptance conditions and integrated into the game arena.  Solving the game amounts to computing the set of winning states---those from which the controller can enforce the objective regardless of the uncontrolled system parts---via a fixed-point computation~\cite{Maler95,Pnueli89}.  
If the initial state $q_0$ lies in this set, a winning strategy exists; otherwise, the specification is unrealisable.

A winning strategy is a mapping
\smaller{$\pi: Q \rightarrow (\Sigma_c \times \mathbb{R}_{\ge 0})$}
assigning to each reachable state a controllable action and delay such that the successors remain within the winning region.  
This strategy constitutes a correct-by-construction controller that guarantees satisfaction of the specified objectives.  
In practice, we employ the \textsc{Uppaal~Tiga} tool~\cite{Behrmann07} for synthesis and validation.
The strategy guarantees that if the human behaves as captured by \HM, the automation achieves its  objectives. 
Since we coarsely abstracted the human, we validate and adapt the automation, if necessary.  
A successful synthesis yields an adaptive design~\cite{landau2011} that explicitly accounts for the learned human model. 

\subsection{Refining the Human Model}
\label{Sec:Refinement}
\begin{figure}[t]
    \centering
    \includegraphics[width=.8\textwidth]{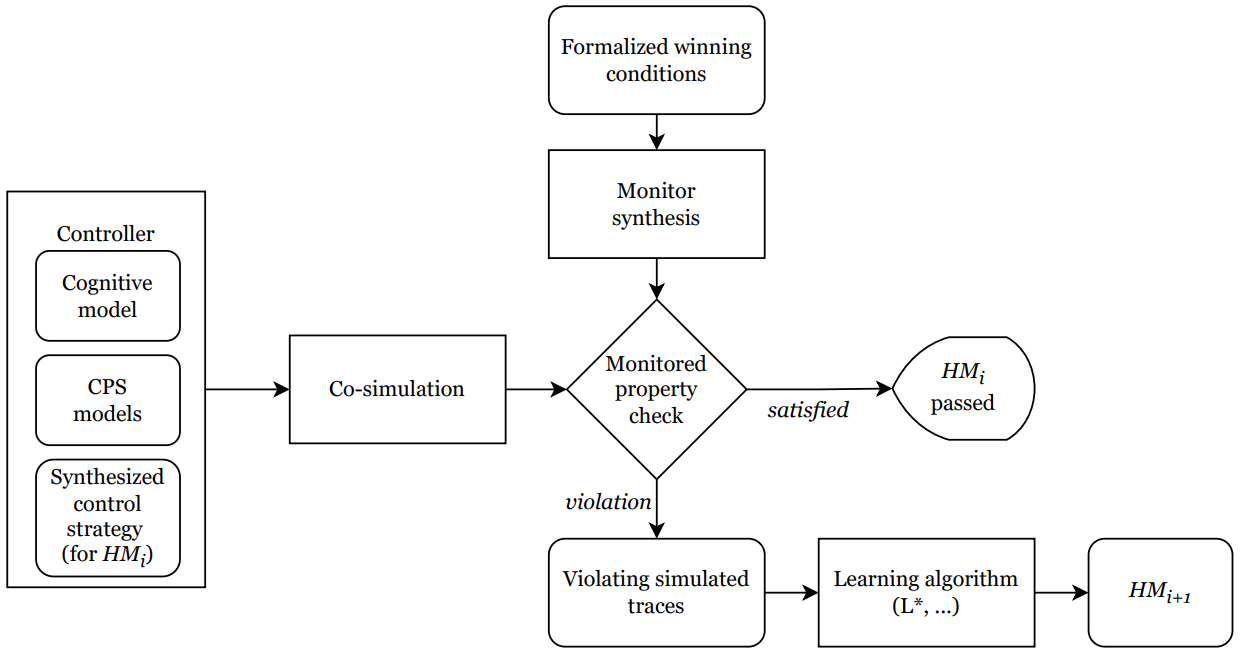}
    \caption{Framework to refine the learned human model \HM.}
    \label{FWRefine}
\end{figure}
Following synthesis, the correctness of the derived automation strategy is evaluated through co-simulation with the full human model.
As illustrated in Fig. \ref{FWRefine} our framework systematically validates the synthesised automation strategies: The full \ACTR model, the environment and the synthesised controller are composed and monitored to verify whether the control objectives are satisfied. 
Failure to meet these objectives indicates that the learned model \HM\ may not yet capture sufficient behavioural fidelity.
Since \HM is inferred from a finite set of coarse traces, its abstraction may omit behaviours relevant to the satisfaction of the control objectives.  

To address this limitation, the framework employs iterative refinement (Fig. \ref{FWRefine}), focusing on traces that lead to specification violations.
The traces are fed back into the learning process to yield an updated model~\HMii{} and a revised strategy~\CSii.  
Each iteration thus extends the behavioural coverage of the human model and enhances the robustness of the synthesised controller.

Although convergence of this loop is not formally guaranteed, termination can be enforced through predefined criteria such as reaching a performance threshold or achieving stability across successive iterations.
Future work will address the formulation of formal termination conditions and the integration of the full refinement loop into case study evaluations, as this mechanism was not yet applied in the present study.
As empirical behavioural data become available, additional refinement can be performed through parameter adaptation, improving the generalisability of the learned model across operator profiles.

\subsection{Adapting the Automation Design Variant}
\label{sec:adapt}
If no winning strategy exists for a given design variant, the automation design is revised---typically by expanding the controllable action set---and synthesis is repeated.  
This iterative design exploration continues until a feasible strategy is identified or all design variants have been exhausted.
Each synthesised strategy is validated through co-simulation as described above, completing the design-time loop of model learning, synthesis, and refinement.
This process supports the systematic development of robust shared-control automation that explicitly accounts for human variability and task context.

\section{Preliminary Evaluation through Case Study}
\label{sec:caseStudy}
\normalsize
We conducted a case study on a driving task to evaluate the feasibility of deriving reactive automation strategies based on a learned human model and to assess whether the obtained behavioural abstraction provide sufficient fidelity for safe shared-control. 
The study focuses on \emph{cognitive-band actions} along the activity continuum~\cite{newell90}---decisions unfolding over time scales of a few seconds. 
In the driving context, these correspond to longitudinal control subtasks, such as adjusting acceleration to maintain a safe headway to a lead vehicle. 
The simulation setup consists of a single-lane road with a \emph{lead vehicle} whose velocity varies over time and a \emph{following vehicle} controlled by the driver model.
The objective is to maintain a safe longitudinal distance from the lead vehicle while adapting to its speed fluctuations.
This configuration provides a controlled environment for analysing shared-control strategies.
\subsection{HM -- the Driver Model}

\begin{figure}[t]
    \centering
    \includegraphics[width=.6\textwidth]{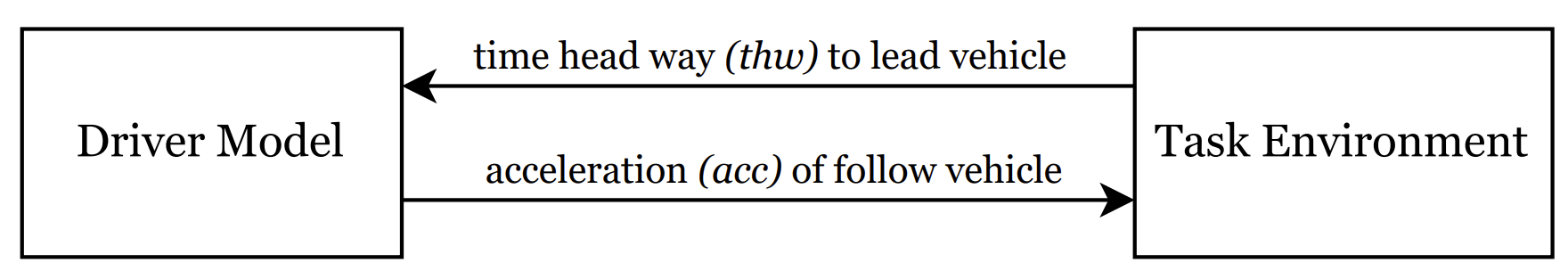}
    \caption{Interaction between driver model and driving environment.}
    \label{fig:DriverEnv1}
\end{figure}

%
We implemented the \emph{driver model} within the \ACTR\ architecture (using the \textit{pyactr} implementation~\cite{pyactr}), integrating goal, declarative, procedural, visual and manual modules.
Its task is to select the longitudinal acceleration of the following vehicle based on the observed time headway~(\thw) to the lead vehicle, \ie the longitudinal gap divided by the speed of the following vehicle.
Specifically, the model evaluates the change in time headway~(\dthw) over elapsed time~(\dt), and computes the acceleration according to an adapted form of the Salvucci driver model \cite{Salvucci06}:
\begin{align*}
    \Delta a = k_1\Delta thw + k_2(thw -thw_{follow})\Delta t.
    \label{EQ:acc}
\end{align*}
Here, $k_1$ penalises abrupt acceleration changes, while $k_2$ drives \thw\ towards the desired value $thw_{follow}$.
At each simulation step,
\begin{enumerate*}[label=(\roman*)]
    \item the environment updates vehicle positions and velocities;
    \item the current \thw\ is computed and passed to the driver model;
    \item the model selects the acceleration for the next step; and
    \item the internal subgoal is updated to ensure behavioural continuity.
\end{enumerate*}
This process establishes a closed-loop interaction between the driver model and its environment (Fig.~\ref{fig:DriverEnv1}).
The simulation generates temporal traces of states and actions $(s_0, a_1, s_1, a_2, ..., s_n)$, where states encode cognitive variables such as buffer contents and goals, and actions correspond to perceptual updates, rule firings, or motor responses.

\paragraph{Scope and rationale.}
The generation of the cognitive model itself is not part of the proposed framework; instead, we modified and used a validated \ACTR-based driver model.
The cognitive mechanisms and empirically grounded representation of decision-making at the cognitive-band level make it a suitable proxy for human control behaviour in the present driving scenario.
This allows the evaluation to focus on the subsequent learning and synthesis steps rather than on model construction.

\subsection{Learning the Human Behaviour Abstraction}
\label{sec:CSabstraction}
We treat the \ACTR\ driver model as a grey-box \emph{system under learning} (\SUL) and infer a finite-state \HM using active automata learning. 
The learned \HM is a finite Mealy machine to represent the automation’s operational view of admissible human input–output behaviour in the given task context.%

\paragraph{What is identified.}
During simulation, a set of tuneable parameters and observable variables are exposed to the learner, grouped into four categories (Table~\ref{Tab:inputs}): 
\begin{enumerate*}[label=(\roman*)]
    \item \emph{Internal parameters}---model-intrinsic variables such as \ACTR\ production rules, $k_1,k_2$, $thw_{follow}$ and reaction latency, which shape the underlying \HM dynamics; 
    \item \emph{Task-specific parameters}---scenario-defining elements such as the lead vehicle profile, initial conditions, and destination, which determine contextual stimuli;
    \item \emph{Simulation parameters}---measured variables and timing quantities (\eg $thw$, $\Delta thw$, $\Delta t$), forming the input alphabet ($\Sigma$) and time context; 
    \item \emph{Outputs}---selected accelerations and relevant cognitive-trace features, constituting the output alphabet ($\Gamma$).
\end{enumerate*}
This mapping enables the learner to capture both observable task behaviour and salient internal reasoning features yielding an \HM that reflects the driver’s decision process in its operational context.
\begin{table}[t]
    \centering
    \scriptsize
    \begin{tabular}{|l|l|l|}
    \hline
        \textbf{Parameter}               & \textbf{Description}                                                   & \textbf{Type of parameter}                 \\ \hline
        \begin{tabular}[c]{@{}l@{}}Declared chunks\\ 
        and procedural rules\end{tabular} 
                                & Internal logic and rule-based system of ACT-R driver model
                                & Internal                                                                                  \\ \hline
        $k_1$, $k_2$            & Coefficients of Eq. 1 tuning driver model behaviour           & Internal         \\ \hline
        $thw\_follow$           & Desired temporal gap the driver aims to maintain               & Internal         \\ \hline
        Reaction latency        & Delay due to ACT-R processing                                 & Internal                      \\ \hline
        Vehicle dynamics        & Update equations for vehicle velocity and position            & Task-specific            \\ \hline
        $destination\_pos$      & Target position of follow vehicle                             & Task-specific                     \\ \hline
        $Init\_conditions$   & Initial positions and velocities of two vehicles              & Task-specific                     \\ \hline
        $lead\_profile(t)$      & Time-varying speed/acceleration profile of the lead vehicle   & Task-specific                     \\ \hline
        $thw$                   & current time headway                        & Simulation            \\ \hline
        $\Delta thw$            & Change in \thw compared to previous step             & Simulation            \\ \hline
        $\Delta t$              & Simulation time step size                                 & Simulation              \\ \hline
        $acc\_follow$           & Acceleration decided by driver  model       & Output                     \\ \hline
        simulation-trace        & Sequence of activated modules, fired production rules and executed actions  & Output                     \\ \hline
    \end{tabular}
    \caption{List of identified parameters.}
    \label{Tab:inputs}
\end{table}
\paragraph{Learning configuration.}
We define the input alphabet $\Sigma$ as discrete environmental stimuli delivered to the driver model, such as quantised levels of \thw. The output alphabet $\Gamma$ comprises tuples combining the decided acceleration and compact representations of the cognitive trace (\eg sequences of fired production rules). 

Each input stimulus triggers internal deliberation that culminates in an output action; the resulting input–output observations populate the L* observation table, where prefixes form the rows, suffixes the columns, and output tuples the cell entries.
An illustrative excerpt is provided in Table~\ref{Tab:OT}.
For instance, the last row and first column indicate that, given the input sequence $(1,1,2,1)$ and query $1$, the learner receives the output \textit{(('attend','read','encode','n\_ret'),-2)}. This corresponds to the sequential firing of the four production rules mentioned, whose combined execution results in a decided acceleration of
$-2$.
\begin{table}[b]
    \centering
    \scriptsize{
    \begin{tabular}{|l|l|l|l|}
        \hline
        Prefixes/E set              & (1,)                                                  & (2,)                                                  \\ \hline
        ()                          & (('attend','read','encode','retrieve','decide'), 0)    & (('attend','read','encode','retrieve','decide'), 0)    \\ \hline
        (1,)                        & (('attend','read','encode','n\_ret'),0)               & (('attend','read','encode','retrieve','decide'),2)    \\ \hline
        (2,)                        & (('attend','read','encode','retrieve','decide'),-3)   & (('attend','read','encode','n\_ret'),0)               \\ \hline
        (1,2)                       & (('attend','read','encode','retrieve','decide'),2)    & (('attend','read','encode','n\_ret'),2)               \\ \hline
        (2,1)                       & (('attend','read','encode','n\_ret'),-3)              & (('attend','read','encode','retrieve','decide'),-1)   \\ \hline
        (1,2,1)                     & (('attend','read','encode','n\_ret'),2)               & None                                                  \\ \hline
        (2,1,2)                     & None                                                  & (('attend','read','encode','n\_ret'),-1)              \\ \hline
        (1,1,2)                     & (('attend','read','encode','retrieve','decide'),-2)   & (('attend','read','encode','n\_ret'),1)               \\ \hline
        (1,1,2,1)                   & (('attend','read','encode','n\_ret'),-2)              & (('attend','read','encode','retrieve','decide'),0)    \\ \hline
    \end{tabular}}
    \caption{Example of observation table.}
    \label{Tab:OT}
\end{table}
\paragraph{Rational for active learning.}
Active automata learning offers a query-driven, sample-efficient approach for identifying finite-state models \cite{angluin87}, supported by mature methodology and tooling \cite{vaadrager2017, howarPotentials, aalpy22}. 
Unlike purely passive, log-based methods, it can deliberately explore rare or safety-critical behaviours that may not appear in empirical data---an essential property for \HCPS safety analysis.
Refinements such as \TTT \cite{IsbernerTTT14} and \NL \cite{Bollig09} improve query efficiency and model succinctness, while recent studies demonstrate continued applicability in complex software settings \cite{Muskardin24}. We therefore adopt an active setup to derive an \HM that is both behaviourally discriminating and synthesis-ready.

\paragraph{Implementation details.}
The active learning process was implemented using \AALPY~\cite{aalpy22} employing a deterministic L* learner in combination with a \smaller{\texttt{RandomWalkEqOracle}} to approximate equivalence queries. The oracle’s reset probability parameter controlled the frequency of simulation restarts during exploration.
The resulting \HM is a deterministic Mealy machine, in which outputs depend on the current state and the provided input. 
For the configuration with four discretised time-headway levels,
the learned \HM comprised 13 states and 52 transitions.

\subsection{Shared-Control Automaton}

To operationalise the synthesised strategy, we implement the automation component as a \emph{supervisory automaton} that monitors safety-relevant variables and dynamically regulates control authority. 
The automaton acts over the \HM, enforcing safety constraints while maintaining a cooperative shared-control interaction.

The controller monitors the time headway (\thw) and time to collision (\ttc), defined respectively as the ratio of inter-vehicle distance to the follower’s velocity and to their relative velocity.
Both quantities are required to remain above specified thresholds (\eg \ \thw $\geq 1.5$s, \ttc $\geq 2$s). 
At each step, the controller evaluates short-horizon predictions (one or two steps ahead) to assess whether the driver’s selected acceleration might violate these thresholds, indicating increased collision risk. 
Based on this assessment, the controller switches between three modes:
\begin{itemize}
    \item Nominal (no hazard): the driver’s acceleration is accepted without modification;
    \item Advisory (low risk): the controller issues an alert signal (``hint'') prompting the driver model to re-evaluate the current \thw\ before finalising its decision; 
    \item Intervention (high risk): if risk exceeds critical limits, the controller overrides or modifies the driver’s selected acceleration, enforcing emergency braking. 
    This represents a transition from advisory support to direct intervention.
\end{itemize}

Transitions between these modes are guarded by hazard predicates:
\smaller{
\begin{align*}
    RiskWarn = (thw < thw_{warn}) \vee (ttc < ttc_{warn}),\\
    RiskFilter = (thw < thw_{min}) \vee (ttc < ttc_{min}),\\
    SafeNow = (thw \geq thw_{safe} \wedge (ttc \geq ttc_{safe}).
\end{align*}}
\normalsize
This defines a compact three-mode shared-control policy with recovery transitions (\eg \emph{Recover-From-Advisory}, \emph{Recover-From-Intervention}) returning the system to lower modes once hazards resolve.

The resulting automaton provides a concrete realisation of the shared-control paradigm explored in this work: it embodies predictive risk awareness, selective intervention, and preservation of human agency.
In the broader context of \HCPS\ design, it serves as a design blueprint for instantiating formally synthesised strategies as interpretable supervisory control logic.

\subsection{Game-Theoretic Synthesis of Shared-Control}
\label{sec:GameSynthesis}
The final step integrates the learned \HM and the automation’s supervisory logic into a unified game-theoretic synthesis framework. 
The synthesis process formally derives a reactive control strategy that guarantees satisfaction of the specified objectives under all admissible human and environmental behaviours.

The game arena is constructed as the synchronous product of the component automata:
\begin{align*}
    \mathcal{A} = Driver \parallel Lead \parallel Follow \parallel SensorError \parallel SimCtrl \parallel Control.
\end{align*}
Each automaton represents a constituent of the \HCPS, contributing its own states, variables, and transition dynamics.

\begin{itemize}
    \item \emph{Driver}: Derived from the learned Mealy-machine abstraction of the cognitive model and translated into a timed automaton. A Python-based tool extracts states and transitions, while timing guards and invariants encode the 50 ms cognitive-cycle constraint (\cf \ ~Sect.~\ref{sec:human}).
    \item \emph{Lead and Follow}: Represent vehicle dynamics; the lead vehicle follows a predefined motion profile, while the following acceleration is governed by the driver model or overridden by the Control automaton during intervention.
    \item \emph{SensorError}: Introduces stochastic perturbations to the perceived \thw\ to represent sensor and perception noise.
    \item \emph{SimCtrl}: Serves as a global scheduler, maintaining consistent timing semantics across all components.
    \item \emph{Control}: Encodes the supervisory shared-control automaton with three modes—\emph{Nominal}, \emph{Advisory}, and \emph{Intervention}—encoding the automation’s decision logic for issuing hints or enforcing overrides.
\end{itemize}

A global system state of $\mathcal{A}$ is defined as 
\begin{align*}
    q = (l_{Driver}, l_{Lead}, l_{Follow}, l_{SensorError}, l_{SimCtrl}, l_{Control}, v),
\end{align*}
where $l_i$ denotes the active location of component~$i$ and $v$ is the joint valuation of clocks and continuous variables (\eg positions, velocities, accelerations). 
Transitions of the \textit{Control} automaton correspond to controllable actions $\Sigma_c$, while all other components contribute uncontrollable actions $\Sigma_u$, representing human behaviour, environment dynamics, and exogenous disturbances.

\paragraph{Specifications and synthesis.}
The control objectives are formulated as winning conditions expressed in a fragment of \TCTL: 
\begin{itemize}
    \item Safety: the follower must never overtake the lead, $\forall \square (follow\_pos < lead\_pos)$;
    \item Reachability: the follower eventually reaches the destination, $\exists \lozenge (follow\_pos \ge DEST)$;
    \item Minimal intervention: the control automaton remains in Nominal or Advisory unless a hazard is detected $\forall \square ((safe \lor low\text{-}risk) \rightarrow \neg Intervention)$. 
\end{itemize}
These objectives jointly yield the weak-until condition:
\[
\texttt{control}: A[\neg(follow\_pos \ge lead\_pos)\ W\ (follow\_pos \ge DEST)]
\]
requiring that the follower never overtakes the leader until the destination is reached.

We employ the \uppaal\ model checker~\cite{Behrmann07} (v5.0.0) to solve the game. 
If the objectives are satisfiable, the tool synthesises a memoryless strategy
$\pi : Q \rightarrow \Sigma_c$,
mapping each reachable state to a controllable action.
The resulting strategy instantiates the three-mode policy:
\[
  \pi(q) = 
  \begin{cases}
    \textit{Nominal:} & \neg RiskWarn \wedge \neg RiskFilter \Rightarrow follow\_acc = driver\_acc,\\
    \textit{Advisory:} & RiskWarn \wedge \neg RiskFilter \Rightarrow \texttt{hint!},\\
    \textit{Intervention:} & RiskFilter \Rightarrow acc\_floor = -3,~acc\_cap = -1.
  \end{cases}
\]

Simulation traces confirm the intended behaviour:
In \emph{Nominal} mode, the driver retains full authority and the invariant $follow\_pos < lead\_pos$ is preserved. 
Under \emph{Advisory}, the controller issues a \texttt{hint!} message, prompting the driver to reassess the situation. 
\emph{Intervention} occurs only under critical conditions (\eg near-collision under sensor noise), enforcing braking until safety is restored. 
Representative trace segments in Table \ref{exampleFr} demonstrate that the synthesised strategy satisfies the weak-until property, maintains safe distance, ensures progress to the destination, and minimises unnecessary overrides.

This synthesis step demonstrates how the learned human abstraction, supervisory control logic, and safety-game formulation combine to realise a correct-by-construction shared-control policy for \HCPS.

\begin{table}[!ht]
    \centering
    \scriptsize{
    \begin{tabular}{|l|l|l|l|l|l|l|}
    \hline
        lead\_pos & follow\_pos   & \thw  & control\_mode         & driver\_acc   & follow\_acc (applied) & Controller action \\ \hline
        172       & 158           & 3     & Intervention          & 2             & –1 (clamped)          & Brake override    \\ \hline
        180       & 165           & 2     & Nominal → Advisory    & –3            & 1                     & Hint issued       \\ \hline
        188       & 175           & 2     & Nominal               & –1            & –1                    & Safe, no action   \\ \hline
        204       & 182           & 4     & Nominal               & –2            & 1                     & Safe, no action   \\ \hline
        204       & 189           & 0     & Intervention          & 2             & –1                    & Brake override    \\ \hline
        212       & 200           & 1     & Advisory → Intervention   & 0           & –1                    & Brake override      \\ \hline
        220       & 198           & 2     & Nominal               & 0             & 1                     & Safe, no action   \\ \hline
    \end{tabular}}
    \caption{Exemplary fragment of simulation trace supervised by the synthesised strategy.}
    \label{exampleFr}
\end{table}

\section{Related Work}
\label{sec:RW}
\normalsize
While the technical aspects of automation within cyber-physical systems (\CPS) have been extensively investigated, the impact on human operators remains comparatively unexplored \cite{lee2004}. 
Recent efforts toward reliable \HCPS aim to integrate models of human behaviour with \CPS models, 
extending formal analysis and synthesis to encompass human–machine interaction.
Human-in-the-loop control synthesis has been studied, for instance, by \cite{Feng16}, in the context of conditional driving automation (\SAE Level 3), where advisory controllers are synthesised to mediate authority transfer between driver and automation.

Bridging formal methods and cognitive modelling remains at an early stage.
Most existing work adopts automata- or Markov-based abstractions of human behaviour to enable verification or strategy synthesis in human–\CPS interaction (\eg \cite{Damm19,sadig16}).
While computationally tractable, these abstractions lack the psychological grounding of cognitive architectures, such as \ACTR \cite{Anderson04}, \smaller{SOAR} \cite{rosenbloom93}, or \CASC by \cite{wortelen13}, which explicitly model perception, decision-making, and learning processes.
Incorporating such architectures into formal frameworks represents a promising but still underdeveloped direction toward cognitively grounded and explainable synthesis of shared-control automation.

Initial steps towards bridging this gap have been taken by \cite{Langenfeld19} and \cite{Gall18}. Both groups pioneered translations of fragments of cognitive architectures into formal models, thereby providing effective reductions of correctness problems of cognitive architectures into model-checking or constraint-solving problems for which effective tool support exists.
Langenfeld et al. \cite{Langenfeld19} mapped \ACTR mechanisms into network of timed automata \cite{Alur94}, enabling automated defect analysis of programmed models. 
While precise, such encoding often leads to significant computational costs when applied to control synthesis.
Current synthesis algorithms do not appear capable of handling timed automata translations of full-fledged \ACTR models, due to both computational scalability limitations and the lack of expressiveness for probabilistic transition selection inherent to \ACTR models. 
The optimized encoding may alleviate the scalability issues, but the application of less heavy-weight strategy synthesis algorithms like reinforcement learning trade off scalability with weaker formal guarantees. 

The cognitive architectures have also served as proxies for human behaviour, particularly in the design of driver-assistance systems (\eg \cite{wortelen13}), and automated risk analysis (\eg \cite{puch13}). However, their potential for automatic synthesis of reactive strategies remains largely unexplored. 
Such a procedure would allow automation to anticipate human behaviour to generate a strategy to fulfil the technical subsystems objectives.
It follows model-predictive control paradigm to account for the expected states and behaviour of human as reflected in the cognitive model.

One of the key challenges is state estimation.
As discussed in our previous work, the existing control synthesis approaches relying on the synthesis methodology of \cite{Maler95} assume full observability of the underlying timed game. 
The strategy is synthesised depending on the entire vector of cognitive states which is unrealistic.
In practice, human cognitive states are largely not observable and can only be inferred indirectly to a limited extent through correlations with (often weak) neurophysiological signals or behavioural cues. 
Practical controllers must therefore remain robust under partial observability, reconstructing latent human states from measurable indicators while tolerating uncertainty.

Approaches addressing scalability and limited observability include abstraction-based methods.
Ehler \etal \cite{Ehlers10} proposed abstract timed games 
that group behaviourally equivalent states with respect to the analysed properties to mitigate state-space explosion.
The key idea is to construct high-level representation of relevant behaviour and iterative refinement to increase accuracy and preserve correctness. 
Similarly, \cite{Johnsen25} integrated automata learning and statistical model checking to derive user-centred strategies in interactive systems..
Their case study on music streaming service demonstrates how learned user models can be used to guide interaction strategies without restricting the user's freedom of choice.
The authors of \cite{vonBerg25} developed probabilistic models of human driver behaviour using a hierarchical learning approach that incorporates known system modes. 
Their model captures highway driving responses to contextual factors such as speed limits and curvature, accurately reproducing observed patterns and generating realistic simulation data.
However, these models remain descriptive rather than explanatory.
Our approach models behaviour through perception, decision-making, and action processes grounded in psychological theory. 
Cognitive architectures offer explainability and enabling analysis of why drivers act as they do, though they may be less precise in continuous control reproduction.
For evaluating shared-control paradigms in \HCPS, such cognitively grounded models are preferable to purely data-driven approaches, as they capture goals and situational awareness.

Our implemented case study illustrates the feasibility of combining human-behaviour learning and reactive synthesis within a shared-control scenario. 
While the results are encouraging, several aspects of the framework merit further reflection regarding methodological choices, assumptions, and applicability to larger systems. 
We therefore discuss these points in more detail in the following section.

\section{Discussion and Future Work}
\label{sec:discussion}
The proposed framework demonstrates how active automata learning and reactive synthesis can be jointly applied to analyse shared-control in \HCPS. 
While the feasibility study confirms the conceptual soundness of the approach, several aspects deserve further refinement, and extension, particularly regarding abstraction design, timing representation, and empirical validation.

\paragraph{Choice of learning algorithm.}
We adopted Angluin’s L* algorithm owing to its simplicity, transparency, and well-understood convergence guarantees. Although more recent variants such as \TTT or \NL improve query efficiency, they remain conceptually based on L* and are already supported by existing learning libraries such as \textsc{Aalpy}. The framework can therefore incorporate these variants without conceptual modification.

\paragraph{Timing and observability assumptions.}
The learned Mealy machine (\HM) captures the discrete decision logic of the \ACTR model, whereas timing aspects are currently introduced through manual annotation. 
Transitions are enriched with timing parameters derived from \ACTR’s internal processing cycle and observed reaction latencies, ensuring consistency with the cognitive architecture’s temporal dynamics. 
Future work will investigate automatic inference of timing parameters, for example through timed-model learning or statistical estimation of timing distributions from simulation traces.

\paragraph{Abstraction design and human-state observation.}
An important consideration concerns how the full cognitive model is abstracted into a form usable for synthesis. 
The abstract human model (\HM) represents the controller’s belief about the operator’s behaviour; its construction must therefore align with the observability of human states.
In practice, the controller continually compares predicted and observed human actions to assess whether its abstraction remains accurate. 
This implies concrete requirements for sensor design and perceptual inference, potentially involving physiological and behavioural indicators such as facial expressions or gaze direction.
The feasible level of abstraction must thus be co-designed with the sensing modalities and computational resources available.
Future work will explore systematic methods for linking abstraction refinement to observability, enabling the automation to adapt its predictive model using measurable human cues.

\paragraph{Refinement of the human model.}
In the current framework, refinement of the human model is triggered by specification violations detected during co-simulation. 
Although this iterative process improves behavioural coverage, formal termination criteria remain to be defined.
Future work will address the formalisation of convergence criteria, potentially using stability measures or bounded performance metrics. 
Moreover, probabilistic or non-deterministic learning algorithms could capture inter-individual variability and uncertainty, strengthening the robustness of the learned abstractions.

\paragraph{Design variants and levels of automation.}
The synthesis process supports systematic exploration of automation levels and design variants. 
Extending this analysis to more complex interaction settings, such as multimodal human–machine interfaces, cooperative robotics, or higher \SAE automation levels, will enable quantitative assessment of trade-offs between safety guarantees, computational scalability, and human engagement. 
A promising direction is the comparison of design variants under equivalent formal objectives to identify architectures that maximise safety while preserving human agency. 

\paragraph{Integration and empirical validation.}
Integration of the complete toolchain, from cognitive simulation (\ACTR) through model learning (\textsc{Aalpy}) to synthesis (\textsc{Uppaal}), remains a technical challenge. 
Automating this pipeline would enable large-scale evaluation across domains and facilitate reproducibility. 
Ultimately, empirical human-in-the-loop validation is required to assess how well the learned abstractions generalise to real human behaviour and whether the synthesised strategies remain effective under real variability.

\paragraph{Outlook.}
Overall, the framework establishes a principled foundation for bridging cognitive modelling and formal synthesis in shared-control systems. 
Future extensions will aim to close the loop between abstraction, observability, and synthesis—allowing the automation to refine its human model dynamically as behavioural evidence accumulates. By grounding abstraction refinement in measurable human states, the framework may evolve towards adaptive, co-regulative automation that maintains safety while remaining sensitive to human intent.

\section{Conclusion}
\label{sec:conclu}
This work demonstrates that combining automata learning of cognitive models with reactive synthesis provides a rigorous foundation for adaptive automation in \HCPS. 
By learning finite-state abstractions of cognitive architectures and integrating them with cyber-physical components in a game-theoretic synthesis framework, the proposed approach enables systematic, correct-by-construction reasoning about shared control and safety in dynamic human–automation interaction. A key contribution is the iterative learning–synthesis–validation loop, which refines both the human abstraction and the control strategy through counterexamples obtained from co-simulation, thereby establishing a principled link between cognitive modelling and formal verification for realising verifiable, adaptive shared control.
\newpage

\bibliographystyle{eptcs}
\bibliography{citations.bib}
\end{document}